\documentclass[amsmath,amssymb,floatfix,prl,epsf,12pt]{article}

\usepackage{graphics}
\usepackage{amsmath,amssymb,bm,graphicx,url,epsfig}
\usepackage[active]{srcltx}
\usepackage[usenames,dvipsnames]{xcolor}

\usepackage{bm}

\newcommand{\rf}[1]{(\ref{#1})}
\newcommand{\beq}{\begin{equation}}

\newcommand{\eeq}{\end{equation}}
\newcommand{\bea}{\begin{eqnarray}}
\newcommand{\eea}{\end{eqnarray}}

\newcommand{\e}{\mbox{e}}
\renewcommand{\d}{\mbox{d}}

\newcommand{\Lam}{\Lambda}
\renewcommand{\b}{\beta}
\renewcommand{\a}{\alpha}

\newcommand{\m}{\mu}

\newcommand{\del}{\delta}

\newcommand{\oh}{\frac{1}{2}}

\newcommand{\dg}{\dagger}

\newcommand{\tr}{\mathrm{tr}\,}

\newcommand{\no}{\nonumber}
\newcommand{\nn}{\no\\}
\newcommand{\non}{\nonumber \\}

\begin{document}

\begin{center}
\vspace{24pt}
{ \large \bf 
A matrix model for strings beyond the c=1 barrier: \\
the spin-s Heisenberg model on  random surfaces. 
}

\begin{center}

\vspace{18pt}


{\sl J. Ambj\o rn}$\,^{a,b}$,  {\sl Sh. Khachatryan}$\,^{c}$
and {\sl A. Sedrakyan}$\,^{c,d}$

\vspace{18pt}
{\footnotesize

$^a$~The Niels Bohr Institute, Copenhagen University\\
Blegdamsvej 17, DK-2100 Copenhagen, Denmark.\\
{ email: ambjorn@nbi.dk}\\

\vspace{10pt}

$^b$~Institute for Mathematics, Astrophysics and Particle Physics (IMAPP)\\ 
Radbaud University Nijmegen \\ 
Heyendaalseweg 135, 6525 AJ, Nijmegen, The Netherlands

\vspace{10pt}

$^c$~Yerevan Physics Institute\\
Br. Alikhanyan str 2, Yerevan-36, Armenia.\\

\vspace{10pt}

$^d$~International Institute for Physics\\
Natal, Brazil.\\
{ email: sedrak@nbi.dk}\\

}
\end{center}

\vspace{48pt}

\end{center}


\begin{center}
{\bf Abstract}
\end{center}
We consider a spin-s Heisenberg model 
coupled to two-dimensional quantum gravity.
We quantize the model using  the Feynman path integral, 
summing over all possible two-dimensional geometries and 
spin configurations. We regularize this path integral 
by starting with the R-matrices defining the spin-s Heisenberg
model on a regular 2d Manhattan lattice.   
2d quantum gravity  is included by defining the R-matrices 
on random Manhattan lattices and summing over these,
in the same way as one sums over 2d geometries using
random triangulations in non-critical string theory.
We formulate a random matrix model where the
partition function reproduces the annealed average of 
the spin-s Heisenberg model over all random Manhattan lattices.
A technique is  presented which reduces the random matrix integration
in partition function to an integration over their eigenvalues.

\newpage

\section{Introduction}
\label{intro}

Non-critical string theory was introduced 
by Polyakov, and one motivation was to describe quark confinement
in QCD  \cite{Polyakov-1981}. The idea was that in order to 
have a string theory away from the critical dimensions
(26 for bosonic string and 10 for the superstring)
one needs to take into account the conformal mode induced by the interaction
between  matter fields on a 2d surface and the intrinsic geometry
of the surface. In this approach the study of non-critical
string theory becomes equivalent to the study of two-dimensional
quantum gravity  coupled to certain conformal matter fields.
The 2d geometry part of this coupled theory, which is the 
theory of the conformal mode, is denoted  quantum Liouville field theory,
and it is determined by the conformal anomaly.
Using a conformal field theory bootstrap approach 
\cite{KPZ, David-2, Distler} it is possible to solve 
analytically  the 2d quantum gravity theory coupled 
to conformal field theories with central charge $c$, but only  
when $c\leq 1$. The critical exponents of the conformal field 
theories are changed from their values in flat spacetime to 
the so-called KPZ-values when $c \leq 1$. Similarly, the 
quantum values of certain geometric observables are 
differ from what one naively would expect for a 
smooth geometry, and also dependent on the central charge,
showing the backreaction of matter on geometry 
\cite{kawai,watabiki,aw,ajw,d-2}.  
The origin of the so-called $c=1$ barrier is not  yet very clear,
but it has prevented us from using the bootstrap solution 
for a bosonic string in four dimensions,
which is the situation one would expect to be relevant for 
the QCD string.   

The wish to understand and to define rigorously non-critical string theory
also led to a lattice formulation, presenting the
two-dimensional random surfaces appearing in the string path
integral as a sum over triangulated piecewise linear surfaces
\cite{Ambjorn, Kazakov, David}. The sum over ``random triangulations''
(or ``dynamical triangulations'' (DT))
can be represented by matrix integrals \cite{Migdal, Kazakov-3},
and these matrix integrals can be modeled such that they 
represent the combined theory of random surfaces and the 
matter content living on these random surfaces. A typical 
example is to define an Ising spin model on a random 
triangulation the same way as one defines the Ising 
model on a regular triangulation. The combined sum 
over random triangulations and Ising spins can be represented 
by a certain (two-)matrix model \cite{kazakov-ising}. 
The combined system (the annealed
average) has a critical point which represents a $c=1/2$ conformal
field theory coupled to 2d quantum gravity. Similarly, one has 
matrix models describing a (p,q) rational conformal field
theory coupled to 2d quantum gravity and the critical 
exponents of these annealed statistical models can be calculated
and they agree with the corresponding KPZ exponents (see \cite{ginsparg,dz}
for  reviews).
However, this class of  matrix models never allowed a useful 
representation of $c >1$ conformal field theories coupled to 
2d quantum gravity. 
 
 The main goal of the present article is to  define 
a matrix model which will
generate a partition function of matter fields with central charge $c>1$
on random surfaces. In a previous paper \cite{AS-1} we have defined 
the XXZ Heisenberg model with spin 1/2 on a so-called 
random Manhattan lattices (RML)
and have shown how to associate a matrix model with  such 
a model. In particular we showed how to perform the integration
over the angular part $U$ of the random matrices $M$ in the 
decomposition $M=U M_d U^+$, where $M_d$ is diagonal. Such an 
integration is crucial when it comes to solving the 
matrix models encountered so far in the context of 2d gravity. 

In the construction above   
a new class of random lattices enters, the random Manhattan lattices.
They arise in a natural way in several situations. 
\begin{figure}[t]
\center
\centerline{\includegraphics[width=75mm,angle=0,clip]{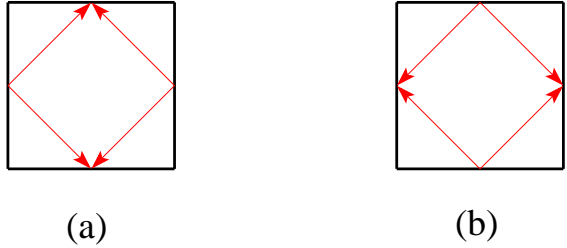}}
\caption{Assignment of arrows to dual lattice}
\label{fig2}
\end{figure}
They appeared in  \cite{Kavalov-1987, AS-1999} 
as  random surfaces embedded in a regular three-dimensional lattice.
Consider the midpoints of the links of a lattice surface embedded in 
a 3d lattice, and join them by new, dual links with
arrows, as it is  shown in Fig.\ref{fig2}. We obtain in this way a
2d-lattice with arrows, 
which will have a Manhattan structure, i.e.\ 
any two neighbour lines will have arrows with opposite directions.
An example is presented in Fig.\ref{fig1}. Any surface in 
3d regular lattice will have as  its dual a Manhattan lattice.

\begin{figure}[h]
\centerline{\includegraphics[width=75mm,angle=0,clip]{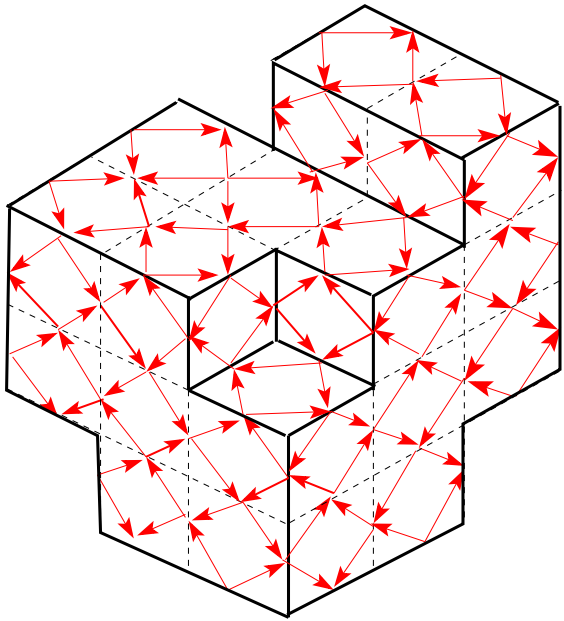}}
\caption{A random surface on the 3d cubic lattice and the 
construction of the dual lattice surface, as described in the text.}
\label{fig1}
\end{figure}

A second way of obtaining a RML is by starting from
oriented double line graphs, like the ones
introduced by 't Hooft (see Fig.\ \ref{4}), 
and then modify the double line propagator like
shown in Fig.\ \ref{fig3}.

\begin{figure}[h]
\center
\centerline{\includegraphics[width=95mm,angle=0,clip]{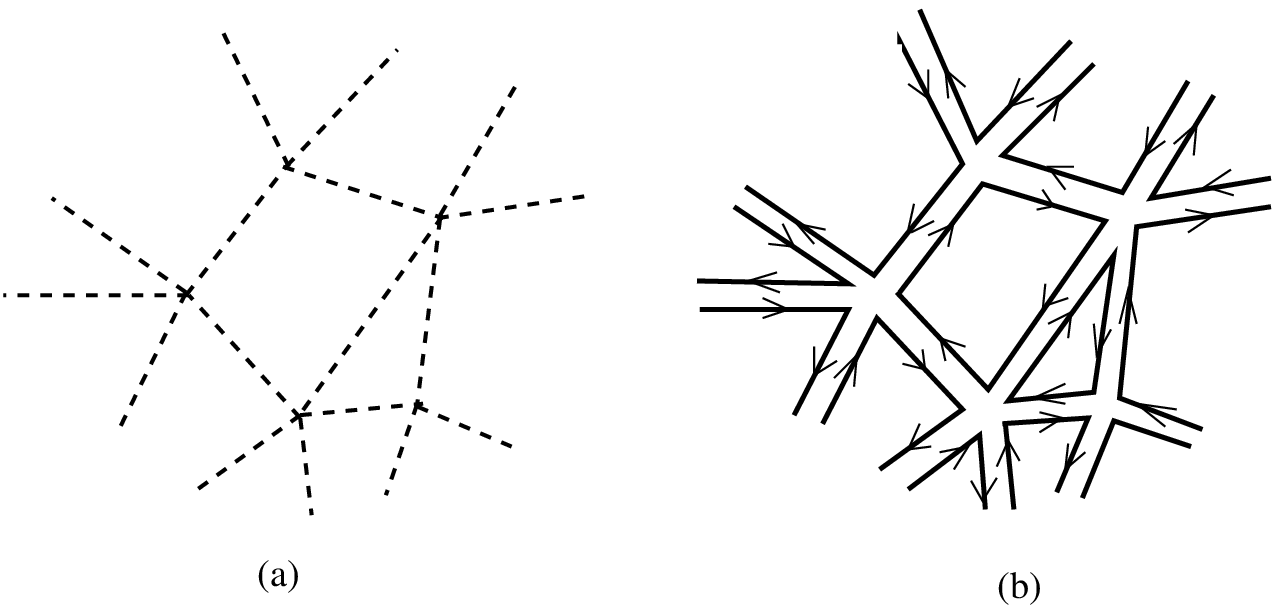}}
\caption{The double line graphs as introduced in gauge theories 
by 't Hooft.}
\label{fig4}
\end{figure}

One is led to such lattices by studying the 
random surface representation of the 3d Ising model  
on a regular 3d lattice \cite{Kavalov-1987, AS-1999},
and via the study of random network of scatterings defined
by S-matrix \cite{Khachatryan-2009, Khachatryan-2010}.
Random networks led to the
idea that an $R$-matrix could be associated to  a random Manhattan
lattice, and we will consider how to couple in general
a matter system defined by an $R$-matrix to a random lattice.
By summing over the random surfaces (i.e. taking the annealed average)
we thus introduce a coupling between the integrable model and two-dimensional
quantum gravity.

The necessity of introduction of RML appears also in studies of plateau-plateau transitions 
in quantum Hall effect (QHE). It appeared, that devoted to plateau transitions in QHE Chalker-Coddington model
 \cite{CC-1988}, which can be formulated as a  network with R-matrices of XX-model on regular Manhattan lattice 
\cite{Sedrakyan-2001}, does not produce  localization length index  compatible
with experiment \cite{Slevin, Amado,  Beenakker, Obuse,  Nuding}. One of possible solution
of the problem is the introduction of the randomness of the network based on RML.

More precisely we start with an integrable model on a 2d square lattice,
assuming we know the R-matrix. We then show that the same R-matrix
can be used on a  random Manhattan lattice.  
On the RML (see Fig.2, Fig.4) the links have fixed arrows which
indicate the allowed  fermion hopping.
No hopping is allowed in directions opposite to arrows.
The summation over the RMLs can be performed by a certain matrix integral
related to the R-matrix. This matrix integral is  somewhat different
from the the conventional matrix integrals used to describe conformal
field theories with $c <1$ coupled to 2d quantum gravity.  This give 
us the hope that  one can penetrate the $c=1$ barrier. Below we present
the construction of a matrix model which will reproduce the partition function
of the spin-s Heisenberg model \cite{Babujian-84} formulated on RML.
The central charge $c$ of excitations in this model 
equal to  $\frac{3 s}{s+1}$, and thus $c>1$ for $s \geq 1$.

\begin{figure}[t]
\center
\centerline{\includegraphics[width=75mm,angle=0,clip]{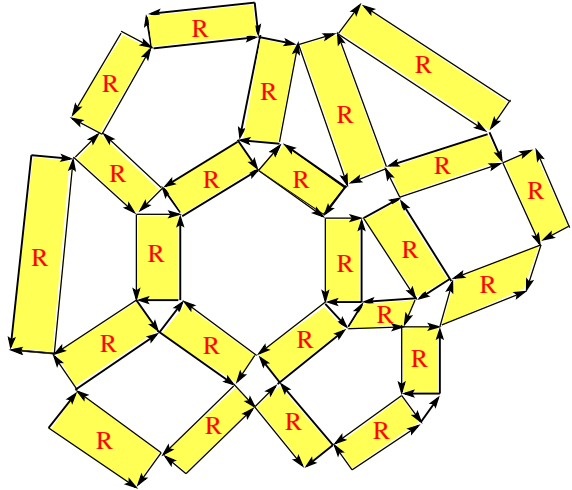}}
\caption{A random Manhattan lattice created from a double line 
graph.}
\label{fig3}
\end{figure}

\section{The matrix model for integrable spin-s Heisenberg model on random surface }

The  $R$-matrix of the spin-s Heisenberg  model is given 
by the following expression \cite{Babujian-84}:
\bea
\label{Rs}
R=-\sum_{j=0}^{2s} \prod_{k=1}^{j}\frac{\lambda-k \eta}{\lambda+k \eta}P^j,
\eea
where 
\bea
\label{Pj}
P^j=\prod_{l=0, l\neq j}^{2 s}\frac{ \vec{S}\otimes\vec{S}-x_l \; 1\otimes 1}{x_j-x_l},\;\; x_j=\frac{1}{2}[j(j+1)-2 s(s+1)]
\eea
are projectors of the product of two spin-s states on an 
irreducible j-spin state.
Here $\lambda$ is a spectral parameter, while $\eta$ is a parameter 
related to the classical R-matrix: when expanding $R$ in powers of
$\eta$ the linear term in the expansion  gives precisely the 
classical R-matrix.

\begin{figure}[ht]
\center
\centerline{\includegraphics[width=95mm,angle=0,clip]{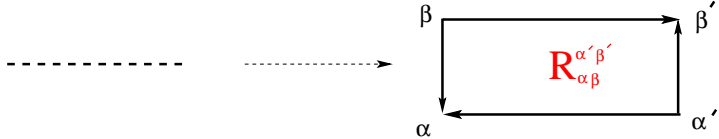}}
\caption{Index assignment of the $R$-matrix}
\label{fig5}
\end{figure}

We now attach it to the squares of
the RML (Fig.\ref{fig3}) with the index assignment shown 
in Fig.\ \ref{fig5}. Two neighbouring
squares will share one of indices, and the same is thus the case for
the corresponding $R$-matrices, and a summation over values of the indices
are understood, resulting in a matrix-like multiplication of $R$-matrices.

To a RML $\Omega$ we now associate the partition function
\beq\label{1}
Z(\Omega) =  \prod_{R\in \Omega} \check{R},
\eeq
where the summation over indices is dictated by the lattice.
Our final partition function is defined by summing over all
possible (connected) lattices $\Omega$:
\beq\label{2}
Z = \sum_\Omega Z(\Omega) \, \e^{-\mu |\Omega|}
\eeq
where $\mu$ is a ``cosmological'' constant which monitors
the typical size $|\Omega|$ of the lattice $\Omega$. 
As long as we restrict the topology of lattices $\Omega$ entering in the 
sum \rf{2}, there will exist a critical $\mu_c$ such that the sum in \rf{2}
is convergent for $\mu > \mu_c$ and divergent for $\mu < \mu_c$.
We will be interested in a limit where the average value 
of $\Omega$ becomes infinite, and this limit is obtained
when $\mu$ approaches $\mu_c$ from above.
The summation over the elements in $\Omega$, i.e.\ the summation over
a certain set of random 2d lattices, is a regularized version of
the sum over 2d geometries precisely in the same way as in
ordinary DT.

In order to represent the spin-s Heisenberg model 
with partition function (\ref{2}), i.e.\ defined on an ensemble of RMLs, 
as a matrix model which
at the same time will offer us a topological expansion of the
surfaces spanned by the oriented ribbon graphs considered above,
we consider the set of $(2s+1)N\times (2s+1)N$ Hermitian matrices. 
We label the entries
of the matrices as $M_{\a\b,ij}$, where $\a,\b$ takes 
values $1,\cdots 2 s+1 $  and
$i,j$ takes values $1,\ldots,N$. 
The $\a,\b$ indices refer to the spin-s Heisenberg
model, while the $i,j$ indices will be used to monitor the topological
expansion. Hermitian  matrix can
be diagonalized by a unitary transformation, i.e. for a given
 matrix $M$ there exists a decomposition
\beq\label{3}
M  = U M^{(d)} U^\dg
\eeq
where $U$ is a unitary $(2s+1)N\times(2 s+1)N$ matrix and
$M^{(d)}$ a diagonal matrix with eigenvalues
$m_{\a,ii}^{(d)}$ which are real  numbers.

Consider now the action
\beq\label{4}
S(M) = M^*_{\a\b,ij} \check{R}^{\a'\b'}_{\a\b} M_{\b'\a',ij} -
\sum_{n=3}^\infty \frac{d_n}{n}\; \tr M^n.
\eeq
We denote the  sum over traces of $M$ as the potential.
The matrix partition function is defined by
\beq\label{4a}
Z = \int \d M \; \e^{-N S(M)}.
\eeq
When one expands the exponential of the potential terms and carries
out the remaining Gaussian integral one will generate all
graphs of the kind discussed above, with the $R$-matrices attached to the
graphs as described. The only difference is that the graphs will be
ordered topologically such that the surfaces associated with the
ribbon graphs appear with a weight $N^\chi$, where $\chi$ is the
Euler characteristics of the surface. If we are only interested in
connected surfaces we should use as the partition function
\beq\label{4b}
F = \log Z.
\eeq
In particular the so-called large $N$
limit, which selects connected surfaces with maximal
$\chi$, will sum over  to
the planar (connected) surfaces  generated by $F$, since these
are the connected surfaces with the largest $\chi$.

The R-matrix (\ref{Rs}) of any spin-s  Heisenberg model can be written
via generators $\lambda_a,\; a=1\cdots (2s+1)^2-1$ 
of the fundamental representation of 
$SU(2 s+1)$ as 
\beq\label{5a}
\check{R}^{\a'\b'}_{\a\b}= \sum_{a=0}^{(2s+1)^2-1} \tilde{I}_a \lambda^{ \a'}_{a;\a} \otimes \lambda^{\b'}_{a;\b}.
\eeq
We  present here the explicit expression of the R-matrix for spin-1 case. 
>From (\ref{Rs}) one can find
\bea
\label{Rs1}
\check{R} = a\; 1\otimes 1 + b \; \vec{S}\otimes \vec{S} +c \; (\vec{S}\otimes \vec{S})^2 ,
\eea
where $S^{\a}_{\b\gamma}=\epsilon_{\a\b\gamma}$ are spin-1 generators of SU(2) and
\bea
\label{Rs2}
a&=&-\frac{1}{3}-\frac{\lambda-\eta}{\lambda+\eta}+\frac{1}{3}\frac{\lambda-\eta}{\lambda+\eta}\frac{\lambda-2\eta}{\lambda+2\eta},\nn
b&=&\frac{1}{2}\frac{\lambda-\eta}{\lambda+\eta} + \frac{1}{2}\frac{\lambda-\eta}{\lambda+\eta}\frac{\lambda-2\eta}{\lambda+2\eta}\\
c&=&\frac{1}{3}+\frac{1}{2} \frac{\lambda-\eta}{\lambda+\eta} + \frac{1}{6}\frac{\lambda-\eta}{\lambda+\eta}\frac{\lambda-2\eta}{\lambda+2\eta}.\no
\eea
It is easy to see that at $\lambda=0$ we have a=1, b=c=0.
One can write the spin-1 $\check{R}$ defined by formula (\ref{Rs1}) 
as  a $9 \times 9$ matrix
\bea
\label{9x9}
\check{R}_1 =\left( \begin{array}{ccccccccc}
a+2 c &0&0&0&b+c&0&0&0&b+c\\
0&a+ c &0&-b&0&0&0&0&0\\
0&0&a+c&0&0&0&-b&0&0\\
0&-b &0&a+c&0&0&0&0&0\\
b+ c &0&0&0&a+2c&0&0&0&b+c\\
0 &0&0&0&0&a+c&0&-b&0\\
0 &0&-b&0&0&0&a+c&0&0\\
0 &0&0&0&0&-b&0&a+c&0\\
b+c &0&0&0&b+c&0&0&0&a+2 c
\end{array}
\right).
\eea
By use of the standard SU(3) generators
\bea
\label{SU3}
\lambda_1=\left(
\begin{array}{ccc}
0&1&0\\
1&0&0\\
0&0&0
\end{array}
\right),\quad
\lambda_2=\left(
\begin{array}{ccc}
0&-i&0\\
i&0&0\\
0&0&0
\end{array}
\right),\quad
\lambda_3=\left(
\begin{array}{ccc}
0&0&1\\
0&0&0\\
1&0&0
\end{array}
\right),\nn
\lambda_4=\left(
\begin{array}{ccc}
0&0&-i\\
0&0&0\\
i&0&0
\end{array}
\right),\quad
\lambda_5=\left(
\begin{array}{ccc}
0&0&0\\
0&0&1\\
0&1&0
\end{array}
\right),\quad
\lambda_6=\left(
\begin{array}{ccc}
0&0&0\\
0&0&-i\\
0&i&0
\end{array}
\right),\\
\lambda_7=\left(
\begin{array}{ccc}
0&0&0\\
0&1&0\\
0&0&-1
\end{array}
\right),\quad
\lambda_8=\frac{1}{\sqrt{3}}\left(
\begin{array}{ccc}
-2&0&0\\
0&1&0\\
0&0&1
\end{array}
\right)\qquad\qquad \no
\eea
a simple calculation shows that indeed the R-matrix (\ref{Rs1}) 
has the  form (\ref{5a}) where
\bea
\label{tilde-I}
\tilde{I}_a=\Big(a+\frac{4 c}{3}, \frac{c}{2}, -\big(b+\frac{c}{2}\big), \frac{c}{2},
 -\big(b+\frac{c}{2}\big), \frac{c}{2}, -\big(b+\frac{c}{2}\big), \frac{c}{3}, \frac{c}{3}\Big)
\eea

Our aim is to decompose the integration over the matrix entries
of $M$ into their radial part $M_d$ and the angular $U$-parameters.
This decomposition is standard and  
the Jacobian is a square of the  Vandermonde determinant.
When we make that decomposition
the potential will only depend on the real eigenvalues $m_{\a,ii}^{(d)}$ and for the
measure we have:
\beq\label{6}
\d M = \d U\;\prod_{\a,i} \d m_{\a,ii}^{(d)} \prod_{\a,i < \b,j}
\Big|(m_{\a,ii}^{(d)}-m_{\b,jj}^{(d)})\Big|^2.
\eeq
However, the problem compared to a standard matrix integral is that
the matrices $U$, introduced by the transformation \rf{3}, will appear
quartic in the action \rf{4}. Thus the $U$-integration does not reduce to
an independent factor, decoupled from the rest. Neither is it
of the Itzykson-Zuber-Charish-Chandra type.

In order to perform the integral over 
$U$ we pass from the expression (\ref{Rs1}),
which is given in the fundamental representation, to a
form where we use the adjoined representation. Let us choose a basis
$t^A$ for Lie algebra of the unitary group $U((2s+1)N)$ in the fundamental
representation. The Hermitian matrix $M$ can also be expended
in this basis:
\beq\label{6a}
M = C_A t^A,~~~~\tr t^At^B = \del^{AB},
\eeq
where the last condition just is a convenient normalization.
For a given $U$ belonging to the fundamental representation of $U((2s+1)N)$
the corresponding matrix in the adjoined representation, $\Lam(U)$,
and the transformation \rf{3} are given by
\beq\label{6b}
\Lam(U)_{AB} = \tr t^A U t^B U^\dg,~~~~C_A = \Lam_{AB} C_B^{(d)},
\eeq
where $C_B^{(d)}$ denotes the coordinates of $M^{(d)}$ in the decomposition
\rf{6a}. The transformation \rf{3} is now
linear in the adjoined matrix $\Lam$ and
the action \rf{4} will be quadratic in $\Lam$. However, we pay of course
a price, namely that the entries of the $((2s+1)N)^2\times ((2s+1)N)^2$
unitary matrix $\Lam$ satisfy more complicated constraints than
those satisfied by the entries of the 
$(2s+1)N \times (2s+1)N$ unitary matrix $U$.
We will deal with the this problem below. First we express
the action \rf{4} in terms of the eigenvalues
$m_{\a,ii}^{(d)}$ and $\Lam$.
\begin{figure}[th]
\center
\centerline{\includegraphics[width=95mm,angle=0,clip]{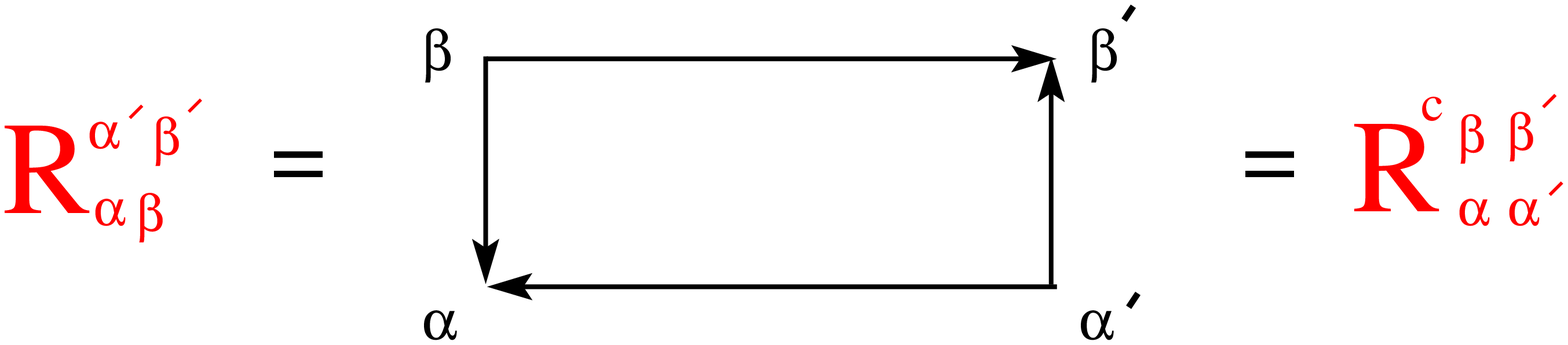}}
\caption{Index assignment of the $R$-matrix and the $R^c$ matrix.}
\label{fig6}
\end{figure}
For this purpose it is convenient to pass from the representation
\rf{5a} of the $R$-matrix to the cross channel  (see Fig.\ \ref{fig6}), which
is achieved by:
\beq\label{7}
\left(\check{R}^c\right)^{\b\b'}_{\a\a'} = \check{R}^{\a'\b'}_{\a\b},
\eeq
which amounts to making a particle-hole transformation $0\leftrightarrow 1$
for the indices $\a'$ and $\b$.  After some algebra one obtains 
for spin-1 case:
\bea
\label{Rc}
\left(\check{R}^c\right)^{\b\b'}_{\a\a'} &=&  (2 b-a)\; 1\otimes 1 + b \; \vec{S}\otimes \vec{S} +(a+c-b) \; (\vec{S}\otimes \vec{S})^2 ,
\eea
For the spin-s case we  write
\beq\label{9}
\left(\check{R}^c\right)^{\b\b'}_{\a\a'} =\sum_{a=0}^{(2s+1)^2-1} I_a \lambda^{ \a'}_{a;\a} \otimes \lambda^{\b'}_{a;\b}.
\eeq
where  $\lambda_0 \equiv 1$, the identity
$(2s+1)\times( 2s+1)$ matrix. For the spin-1 case one has from  (\ref{Rc}):
\bea
\label{10}
I_a = \frac{1}{2}\left( \frac{2(a+2 b+4 c)}{3}, a+c-b,-(a+b+c), a+c-b,-(a+b+c),\right. \nn
\left. a+c-b,-(a+b+c), \frac{2(a+c-b)}{3},\frac{2(a+c-b)}{3}
 \right)
\eea
Using the cross channel $R$-matrix \rf{9} the action \rf{4} becomes:
\beq\label{11}
S= M^*_{\a\b,ij}
\left(\check{R}^c\right)^{\b\b'}_{\a\a'} \del^{i'}_i \del^{j'}_{j}
M_{\b'\a',i'j'} - V(M).
\eeq

Let $\tau^\m$, $\m=1,\ldots,N^2$ denote the generators of the Lie algebra of
$U(N)$, appropriately normalized such that
\beq\label{12}
\del^{i'}_{i} \del^{j'}_{j} = \tau^\m_{ij}\tau^\m_{i'j'}.
\eeq
If we insert \rf{12} into the action \rf{11} we obtain
\beq\label{13}
S= \tr \left( M^\dg\lambda^a \tau^\m\right) I^a
\tr \left(\lambda^a \tau^\m M\right)
\eeq
where we can view $\lambda^a\tau^\m$, $a=0,1,\cdots (2s+1)^2-1$ and $\m=1,\ldots,N^2$ as
the $((2s+1)N)^2$ generators $t^A$ of the Lie algebra of $U((2s+1)N)$.
Formulas \rf{6a} and \rf{6b} with the generators
$\lambda^a\tau^\m$ read
\beq\label{15}
m^{(d)}_{a,\m} = \oh \tr \left(M^{(d)} \lambda^a \tau^\m \right),
\eeq
where only the Cartan generators of $\lambda^a$  and $\tau^\m$ 
give non-zero contributions and
\beq\label{16}
\Lam^{a\m,a'\m'} =
\oh \tr \left(\lambda^a \tau^\m U  \lambda^{a'} \tau^{\m'} U^\dg \right).
\eeq
We now want to use \rf{3} and \rf{15} and \rf{16} to express
the matrix $M$ in the action \rf{13} in terms of $m^{(d)}_{a,\m}$ and
$\Lam^{a\m,a'\m'}$ and we obtain
\beq
S= m^{(d)}_{a'\m'} \Lam^{a\m,a'\m'} I^a \Lam^{a\m,a''\m''}m^{(d)}_{a''\m''}
-V(m^{(d)}_{a\m})
\label{14}
\eeq

It is convenient to choose following basis for generators 
$ t_A,\; A\equiv (a, \mu) \equiv (\a\b, ij)=1\cdots (2 s+1)^2 N^2$ 
of $ U((2s+1) N)$: for
the $(2s+1)N$-dimensional Cartan sub-algebra 
we take $t_{1,1}=1\otimes 1$ for the common phase factor  and   
$t_{\a,i}=\lambda^{\a\a} \otimes (\tau_{ii}- \tau^{i-1,i-1}) ,\; 
\a=1\cdots 2 s+1,  \; i=2\cdots N$ for $(2s+1)(N-1)$ of the 
traceless generators. Finally we choose 
$t_{\a,1}=\lambda^{\a\a}\otimes \tau^{11}-
\lambda^{\a-1,\a-1}\otimes \tau^{N,N},\; \a=2\cdots 2s+1 \ $ 
for the remaining $ 2 s$ traceless generators,  where
\beq\label{17}
\left(\tau^{ij}\right)_{kl} = \del_{ik}\del_{jl}, \;\; \left(\lambda^{\a\b}\right)_{\a'\b'} = \del_{\a\a'}\del_{\b\b'}.
\eeq
For the non-diagonal generators we take 
$\lambda^{\a\b} \tau^{ij} ,\; \a,\b=1\cdots 2s+1, \; i , j=1\cdots N$, 
where either $\a \neq \b $ or  $ i \neq j $.
Then, for this choice of generators and from 
$m^{(d)}_{a,i} = \oh \tr \left(M^{(d)} t_{\a,i}\right) $
 we have
\bea\label{17a}
m^{(d)}_{1,1} &=& \tr M=\sum_{\a=1 \cdots 2s+1; i=1\cdots N} m^{(d)}_{a,ii}\nn
m^{(d)}_{\a,i} &=& m^{(d)}_{\a,ii}-m^{(d)}_{\a,i-1\;i-1},\;\qquad  \a=1 \cdots 2s+1, \;\; \; i=2 \cdots N\\
m^{(d)}_{\a,1} &=& m^{(d)}_{\a,11}-m^{(d)}_{\a-1,NN},\; \qquad \a=2 \cdots 2s+1.
 \label{17b}\no
\label{17c}
\eea
with the rest of elements $m^{(d)}_{a,ij}=0$. It is convenient to impose following order for real eigenvalues
\bea
\label{order}
m^{(d)}_{\xi}=\Big(  m^{(d)}_{1,i}  \cdots m^{(d)}_{\a,i},  m^{(d)}_{\a+1,i}\cdots m^{(d)}_{2s+1,N} \Big),
\eea
then the term $m^{(d)}_{\a,i}$ can be numbered as $m^{(d)}_{(\a-1)N+i}$.

\subsection{The integral over angles: Flag manifolds}

We now change the integration over the unitary matrices $U((2s+1)N)$ in
formula \rf{6}, which are in the fundamental representation, to the
unitary matrices $\Lam((2s+1)N)$ in the adjoined representation.
Rather than using the Haar measure expressed in terms of the $U$-matrices
we should express the Haar measure in terms of the $\Lam$-matrices.

Since Hermitian matrices $M$ can be regarded as  elements in the algebra $u((2s+1)N)$,
the action of $\Lam$, defined by the formula (\ref{6b}) on its
diagonalized form (\ref{17a}), will form an orbit
in the algebra with the basis consisting of all diagonal matrices.
Diagonalized elements of $M$ are invariant under the action of the maximal
abelian (Cartan) subgroup $\otimes U(1)^{(2s+1)N}$of $U((2s+1)N)$.
Therefore the orbits are isomorphic to the factor space
$\frac{U((2s+1)N)}{U(1)\,\otimes \cdots \otimes \,U(1)}$.

Moreover this factor space is isomorphic
to a so-called flag-manifold, defined as follows
(see \cite{Brion, Fuks} and references there):
A single flag is a sequence of nested complex subspaces in a
complex vector space $C_n$
\bea
\label{flag}
\{\varnothing\}=C_0 \subset C_{a_1} \subset \cdots \subset C_{a_k}
\subset C_n=C^n
\eea
with complex dimensions $dim_C C_i=i$.
For a fixed set of integers $(a_1, a_2 \cdots a_k, n)$ the collection of all
flags forms a manifold, which is called the 
flag manifold $F(n_1,n_2,\cdots n_k)$, where  $n_i=a_i-a_{i-1} $.
The manifold $F(1,1,\cdots 1)$ is called a full flag manifold,
the others are partial flag manifolds.
The full flag manifold $F(1,1,\cdots 1)$ is isomorphic
to the orbits of the action of the adjoined
representation of $U((2s+1) N)$ on its algebra
\bea
\label{orbit}
F(1,1,\cdots 1) = \frac{U((2s+1) N)}{\otimes U(1)^{(2s+1)N}}
\eea

The set of $C_{i}$ hyperplanes in $C_{i+1}$ is isomorphic
to the set of complex lines in $C_{i+1}$.
In differential geometry this set is denoted by  $\mathbf{CP}^{i}$ (and
also  as Grassmanians $\mathbf{Gr}(1,i)$) and is called a
complex projective space.  Hence, the complex projective
space is a factor space
\bea
\label{cp}
\mathbf{CP}^{i}= \frac{U(i+1)}{U(i)\otimes U(1)}= \frac{S^{2 i+1}}{U(1)}
\eea
where $S^{2 i+1}$ is a real $2 i+1$ dimensional sphere.

According to description presented above the orbit of
the action of the adjoined representation
of  $U((2s+1) N)$ on the set of normal matrices $M$ is a sequence of
fiber bundles and locally, on suitable
open sets,  the elements of the flag manifold
can be represented as a direct product of projective spaces
(the fibers)
\bea
\label{FB}
\frac{U((2s+1) N)}{\otimes \,U(1)^{(2s+1)N}} 
\simeq \mathbf{CP}^{(2s+1) N-1} \times \mathbf{CP}^{(2s+1) N-2}
\times \cdots \mathbf{CP}^{1}
\eea

In simple words we have the following representation of the orbit:
any diagonalized Hermitian matrix in the
adjoined representation has the following form
\bea
\label{diagonal-form}
M^{(d)}_{a\mu}&=&\Big(\underbrace{m^{(d)}_{2s+1,N},0,\cdots 0}_{2(2s+1) N-1},
\underbrace{m^{(d)}_{2s+1,N-1},0,\cdots 0}_{2(2s+1) N-3},
\cdots \underbrace{m^{(d)}_{\a,i},0,\cdots 0}_{2((\a-1)N+i)-1},\non
& \cdots &\underbrace{m^{(d)}_{1,2},0,0}_{3},m^{(d)}_{1,1}\Big)
\eea
The action of the adjoined representation $\Lam$ on this
$M$ transforms it into the elements of 
$\frac{U((2s+1) N)}{\otimes U(1)^{(2s+1)N}}$
presented in (\ref{FB}) where $\mathbf{CP}^{(\a-1)N+i-1}$
represents image of the part 
$\underbrace{m^{(d)}_{\a,i},0,\cdots 0}_{2((\a-1)N+i)-1} $.

This implies that the measure of our integral
over Hermitian matrices $M$ can be decomposed
into the product of measures of the base space (the diagonal matrices)
and the flag manifold (the fiber)
\bea
\label{measure}
{\cal D}\Lam &=& \prod_{i=1,\;\a=1}^{N,\; 2s+1} d m^{(d)}_{\a,ii} 
\prod_{k=1}^{(2s+1) N-1} {\cal D}[\mathbf{CP}^{k}]\non
&=& \prod_{i=1,\;\a=1}^{N,\; 2s+1} d m^{(d)}_{\a,ii}  
\prod_{k=1}^{(2s+1) N} {\cal D}\Big[\frac{S^{2 k-1}}{S^1}\Big]
\eea

However, since 
our action (\ref{14})  is invariant over $\otimes \, U(1)^{(2s+1)N}$ (one $U(1)$
per marked segment in (\ref{diagonal-form}))
we can extend the integration measure from (\ref{measure}) to
\bea
\label{measure-2}
{\cal D}\Lam =\prod_{i=1,\;\a=1}^{N,\; 2s+1} d m^{(d)}_{\a,ii}  
\prod_{k=1}^{(2s+1) N} {\cal D}\Big[S^{2 k-1}\Big]
\eea
In other words, we suggest that the action of $\Lam$
on the segments $\underbrace{m^{(d)}_{\a,i},0,\cdots 0}_{2((\a-1)N+i)-1} $  in
(\ref{diagonal-form}) form vectors
$m^{(d)}_{\a,i} z_{\a,i}^r,\; (r=1 \cdots 2(\a-1)N+i)-1 $
where the  coordinates $z_{\a,i}^r$ can be considered real parameters belonging 
to the unite spheres $S^{2((\a-1)N+i)-1}$.

In order to write the measure of integration over the
spheres $S^{2k-1}$  we embed them into the Euclidean spaces
$R^{2k}$  and define
\bea
\label{measure-3}
{\cal D}\big[S^{2(\a-1)N+2 i-1}\Big]& =&
\delta\Big(\sum_{r=1}^{2(\a-1)N+2 i} [z_{\a,i}^r]^2-1\Big)\prod_{r=1}^{2(\a-1)N+2 i} dz_{\a,i}^r\\
&=&\int d\lambda_{\a,i}\prod_{r=1}^{2(\a-1)N+2 i-1} dz_{\a,i}^r 
e^{-\lambda^2_{\a,i}(1-\sum_{r=1}^{2(\a-1)N+2 i-1} [z_{\a,i}^r]^2)},\no
\eea
where we have introduced Gaussian integrations over the 
real parameters $\lambda_{\a,i},\; \a=1 \cdots 2s+1,\; i=1, \cdots N $.  These integrations 
reproduce the  factors \\
$ \frac{1}{2 \sqrt{1-\sum_{r=1}^{2(\a-1)N+2 i-1} [z_{a,k}^r]^2}}$ which arise from the 
$\delta$-functions in \rf{measure-3} by  integrations over the  coordinates 
$ z_{\a,i}^{2(\a-1)N+2 i}$. 
We  have omitted coefficients
$\sqrt{\pi}/2$ in front of integrals on the  right hand side of 
the expressions (\ref{measure-3}) since they
unimportant for the partition function.

With this definition of the measure the partition function (\ref{4a})
can be written as
\bea\label{21b}
\int \d M e^{-N S(M)} &=&
\int \prod_{\a=1,\; i=1}^{(2s+1),\; N} \d m^{(d)}_{\a,ii} \d\lambda_{\a,i}\\
&\cdot& \prod_{r=1}^{2(\a-1)N+2 i} dz_{\a,i}^r
 W(m^{(d)}_{\alpha,ii}) e^{-S(m_{\a,i}^{(d)},\lambda_{\a,i},z_{\a,i}^r)},\no
\eea
where $W(m^{(d)}_{\alpha,ii}) =\prod_{\a,i < \b,j}
\Big|(m_{\a,ii}^{(d)}-m_{\b,jj}^{(d)})\Big|^2 $ is the  square of  
the Vandermonde determinant and
\bea
\label{action-2}
S(m_{\a,i}^{(d)},\lambda_{\a,i},z_{\a,i}^r)=
\sum_{\a=1,\;i=1}^{2s+1,\;N} \Big[|m^{(d) }_{\a,i}|^2 
\sum_{r=1}^{s(\a,i)}|z_{\a,i}^{r}|^2 I^r_{\a,i}
+\lambda^2_{\a,i}\big(1-\sum_{r=1}^{s(\a,i)} [z_{\a,i}^r]^2\big)- 
V(m^{(d)}_{\a,ii})\Big].\nn
\eea
Here $s(\a,i)=2(\a-1)N+2 i-1$, while $I^r_{\a,i}$ is defined 
in accordance with expression  (\ref{9})
\bea
\label{Ia}
I=\Big(\overbrace{I_a,I_a, \cdots \underbrace{I^1_{\a,i}, 
\cdots I^{s(\a,i)}_{\a,i}}_{s(\a,i)}, \cdots I_a, I_a}^{N^2}\Big),
\eea
where each row-vector $I_a$ has  $(2 s+1)^2$ elements, 
given in the adjoined representation of U(2s+1).
The total length of $I$  is precisely $(2s+1)^2 N^2$ (as it should be). 
For the spin-1 case the explicit  expression for  $I_a$  
is given by Eq.~(\ref{10}). The $2s+1$ elements  in $I^a$ 
appear  in a specific sequence in $I$ as defined in eq.\ (\ref{Ia}).
However  the partition function is independent of this choice (which 
is just our arbitrary choice) after integration.   
$I^r_{\a,i},\; r=1, \cdots s(\a,i)$ denote the elements
placed in the  position $(\a,i)$  in  (\ref{Ia}).

As one can see we have in the partition function (\ref{action-2})
simple Gaussian integrals over $z_{\a,i}^r$. These can be evaluated
and we are left with integrals over $m^{(d)}_{\a,i}$ and $\lambda_{\a,i}$ only.
It is convenient to rescale the Lagrange multipliers
and introduce  $\tilde{\lambda}_{\a,i}= |m^{(d)}_{\a,ii}|^{-1}\lambda_{\a,i} $.
Then, after performing the Gaussian integrals, we obtain
\bea
\label{27}
Z &=& \int  \prod_{\a=1,\; i=1}^{2s+1,\; N} \d m^{(d)}_{\a,ii}
W(m^{(d)}_{a,ii})  \prod_{\a=1,\; i=1}^{2s+1,\; N}  \frac{1}{|m^{(d)}_{\a,i}|^{s(\a,i)-1}}
\int \prod_{\a=1,\; i=1}^{2s+1,\; N}\d\tilde{\lambda}_{\a,i}\; \nonumber\\
&\cdot&\prod_{\a=1; i=1}^{2s+1, N}e^{- V(m^{(d)}_{a,ii})- |m^{(d)}_{a,i}|^2 \tilde{\lambda}^2_{\a,i}}
 \prod_{\a=1; i=1}^{2s+1, N} Z_{\a,i} (\tilde{\lambda}_{\a,i}).
\eea
where
\bea
\label{Z}
Z_{\a,i} (\tilde{\lambda}_{\a,i})=\prod_{r=1}^{s(\a,i)}
\frac{1}{(I^r_{\a,i}- \tilde{\lambda}^2_{\a,i} )^{1/2}}
\eea

Let us demonstrate that the Gaussian integration  
over the adjoined representation matrices $\Lambda$
in (\ref{27}) correctly reproduces the partition function (\ref{4a}) 
when interaction is absent, i.e.\ $V(M)=0$. In this case the integral
over normal matrices $M$ in the fundamental representation of 
$U((2s+1) N)$ can  easily be performed directly  and the result is:
\bea
\label{Z2}  
Z_{V=0}=\frac{1}{Det[\check{R}_{\a\b}^{\a'\b'}]^{N^2/2}},
\eea
where $\check{R}_{\a\b}^{\a'\b'} $ should be considered as 
a $(2s+1)\times(2s+1)$ matrix.
For spin-1 case it can be calculated directly by use of 
expression (\ref{9x9}) and we get
\bea
\label{Z3} 
Z_{V=0}=\Big[(a+2b+4c)(a+c-b)^5(a+b+c)^3\Big]^{-N^2/2}
\eea

Let us first consider the simple case $N=1$.
In the general setup this corresponds to having the two shortest, 
length 3 and 1 segments in the sequence   (\ref{diagonal-form}).
The Vandermonde determinant cancels the $m$'s 
in the denominator in the expression   (\ref{27}) of the partition function 
and integration over the $m$'s leads to 
\bea
\label{N=1}
Z= \int \prod_{\a=1}^9 \frac{\d\tilde{\lambda}_{\a,1}}{\tilde{\lambda}_{\a,1}}
  \frac{1}{(I^1_{\a,1}- \tilde{\lambda}^2_{\a,1} )^{1/2}}
= \frac{1}{[(a+2b+4c)(a+c-b)^5(a+b+c)^3]^{1/2}}
\eea
provided we place the $\lambda$-poles at zero and the $\lambda$-branch 
cuts at different sides of the real $\lambda$-axis. Here we have used expression (\ref{10})
for $I_{\a,1}^1$.

Now consider a general $N$. For a  generic $i$ segment in 
(\ref{diagonal-form}) we first represent the multipliers 
$|m^{(d)}_{\a,ii}-m^{(d)}_{\b,jj}|$  in the  
Vandermonde determinant as a sum over $m^{(d)}_{\a,i}$:   
For $(\a,i)>(\b,j) $ we can write
\bea
\label{last1}
|m^{(d)}_{\a,ii}-m^{(d)}_{\b,jj}|&=&
|m^{(d)}_{\a,i}+m^{(d)}_{\a,i-1}| \;\;\qquad  for \;\;\quad i>1 \nn 
|m^{(d)}_{\a,11}-m^{(d)}_{\b,jj}|&=&
|m^{(d)}_{\a,1}+ m^{(d)}_{\a-1,N}| \;\qquad  for \;\;\quad i=1 .
\eea 
When using this decomposition in the 
 Vandermonde determinant  product, we observe that only the  
contribution of first terms $ |m^{(d)}_{\a,i}|$ and $ |m^{(d)}_{\a,1}|$
in (\ref{last1}) will cancel all  $m^{(d)}$'s in the denominator of 
(\ref{27}) and thus lead to  a non-zero contribution. By  Cauchy
integration as above we obtain
\bea
\label{N}
Z(V=0)&=& \int \prod_{\a=1;i=1}^{9,N}  
\frac{\d\tilde{\lambda}_{\a,i}}{\tilde{\lambda}_{\a,i}} 
 Z_{\a,i} (\tilde{\lambda}_{\a,i})\nn
&=& \frac{1}{[(a+2b+4c)(a+c-b)^5(a+b+c)^3]^{N^2/2}}.
\eea  
Other terms in the decomposition (\ref{last1}) will result 
in $\tilde{\lambda}_{\a,i},\; \a=1,\cdots 9$ appearing in the denominator
of the integral (\ref{N}) with other powers than one, and 
the integration will give zero for these terms.

\section{Conclusions}
  
We have defined a matrix model which reproduces the partition function of an
integrable spin-s Heisenberg model \cite{Babujian-84} on  random surfaces.
The random surfaces under consideration appear
as random Manhattan lattices, which are dual to random surfaces
embedded in  a $d$ dimensional
regular Euclidean lattice.
This formulation allows us to consider a
new type of non-critical strings with $c >1$,  since   $c=3 s/(s+1)$ for  spin-s Heisenberg model
and it is larger than one for $s\geqslant 1$.

We have shown that  the matrix integral can be reduced to
integrals over the eigenvalues of matrices.
The important ingredient in the integration
over angular parameters of the matrices
(which are usually defined by the Itzykson-Zuber integral)
is the reduction of the problem to an integration 
over unitary matrices in the adjoined representation, 
which is equivalent to flag manifold.
In principle one can now try to apply standard large 
N saddle point methods for solving the resulting integrals 
over eigenvalues.  

The approach presented here  
can be applied to other integrable models where 
one knows the $R$ matrix.


\section{Acknowledgement}
A.S thanks Niels Bohr Institute and Armenian Research Council for 
partial financial support.
JA acknowledges support from the ERC-Advance grant 291092,
``Exploring the Quantum Universe'' (EQU). The authors acknowledge support 
of FNU, the Free Danish Research Council, from the grant 
``quantum gravity and the role of black holes''. Finally this research 
was supported in part by the Perimeter Institute of Theoretical Physics.
Research at Perimeter Institute is supported by the Government of Canada
through Industry Canada and by the Province of Ontario through the 
Ministry of Economic Development \& Innovation.

\end{document}